\documentclass[aps,12pt,superscriptaddress,amsfonts,amssymb,amsmath]{revtex4}

      \usepackage{epsfig}
      \usepackage{makeidx}
         \usepackage{graphics}
      \usepackage{graphicx}
      \usepackage{epstopdf}

\begin{document}

\markboth{M.L. Bertotti \and G. Modanese}
{Microscopic models for the study of taxpayer audit effects}


\title{MICROSCOPIC MODELS\\ 
FOR THE STUDY OF TAXPAYER AUDIT EFFECTS}



\author{Maria Letizia BERTOTTI}

\address{Faculty of Science and Technology, Free University of Bozen-Bolzano, Piazza Universit\`a 5\\
39100 BOLZANO, ITALY, \\
marialetizia.bertotti@unibz.it}

\author{Giovanni MODANESE}

\address{Faculty of Science and Technology, Free University of Bozen-Bolzano, Piazza Universit\`a 5\\
39100 BOLZANO, ITALY\\
giovanni.modanese@unibz.it}

\maketitle


{\bf Abstract -} 
A microscopic dynamic model is here constructed and analyzed, describing the evolution of the income distribution 
in the presence of taxation and redistribution in a society in which also tax evasion and auditing processes occur. 
The focus is on effects of enforcement regimes, characterized by different choices of the audited taxpayer fraction 
and of the penalties imposed to noncompliant individuals. A complex systems perspective is adopted: 
society is considered as a system composed by a large number of heterogeneous individuals. These are divided 
into income classes and may as well have different tax evasion behaviors. The variation in time of the number 
of individuals in each class is described by a system of nonlinear differential equations of the kinetic discretized 
Boltzmann type involving transition probabilities.
\section{Introduction}
\label{intro}

In this paper a microscopic model is constructed and analyzed,
suitable to describe the taxation process in a closed market society in conjunction
with the occurrence of tax evasion and of some audit procedure.
The focus is on effects on the population income distribution caused by
different choices of the audited taxpayer fractions and by
different subsequent penalties imposed to non-compliant individuals.

\smallskip

As a serious, long-standing problem, tax evasion has been the object of numerous studies. 
And being a genuinely interdisciplinary issue, it has been looked at from various viewpoints:
social, economical, behavioral and psychological aspects have been explored
by means of both experimental and theoretical methods.
It is however only during the last decades that novel approaches have been developed, allowing
the inclusion of more realism in the treatment of the problem.
These approaches benefit from a significantly increased computing power
and typically involve agent-based modelling and simulations. Their specificity
is that they provide the possibility of
taking into account the interactions of a large number of heterogeneous individual units.  
From the whole of these interactions, the emergence of some aggregate pattern can be deduced.
Essentially in parallel also alternative approaches to socio-economic questions, which employ concepts and tools 
from statistical mechanics, have been designed by the physics community.
The agent-based methodology and the ``econophysics'' field share the perspective of a {\it {complexity economics}}.
Actually, a clear-cut distinction between the two research lines is not always possible; see \cite{1} for a discussion and a list of references. 

\smallskip

Also in this paper a complex system perspective is adopted. Indeed, the emergence of the aggregate pattern represented by
the income distribution is obtained as the result of a multiplicity of interactions between single units.
These interactions consist of monetary exchanges between pairs of individuals, payment of taxes, redistribution of the tax revenue,
and payment of penalties.
The approach we follow differs from the agent-based methods and from others in the econophysics literature.
It is characterized by its mathematically founded nature. We build on our previous work 
and introduce a modified version of the model discussed in \cite{2} and \cite{3}
(see also \cite{4} and \cite{5}).
Very briefly, we deal with society as with a system composed by a large number of heterogeneous individuals
divided into classes which are distinguished by their average income. The mechanism according to which the individuals exchange money,
pay taxes and benefit of some revenue redistribution (leaving the total wealth unchanged)
is described through a system of nonlinear ordinary differential equations of the kinetic-discretized Boltzmann type,
involving transition probabilities. The evolution equations are as many as the income classes.
Each one describes the variation in time of the number of individuals in a class.
The novel aspect  in this paper, as discussed in detail in Section \ref{model structure} below,  
is that we also incorporate in the equations some terms accounting for
income tax audit process as well as for related penalty payments
from non-compliant audited individuals.
Looking at numerical solutions, we observe that in the long run they all tend to an equilibrium.
This macro-level equilibrium represents the observable income distribution.
We compare scenarios corresponding to cases in which different 
percentages of individuals are being audited and non-compliant audited individuals are differently fined.
In all these cases, we evaluate the tax revenue and the Gini index.
Our interest is to try and see whether it is more convenient to increase the severity of punishment or the number of audits.
Of course, the model is somehow schematic and naive.
Yet it allows explorative simulations and provides answers not obtainable by other means.
We think it could help to better understand the consequences of different policies.

\smallskip

In the classical economic literature the combined effect of tax evasion, audits and payment of fines has been studied in detail 
in the context of representative-agent models (see e.g. the recent papers \cite{6} and \cite{7}. 
In this approach a rational agent, supposed to represent the behavior of the average individual, 
optimizes her/his choices considering several financial factors 
(capital, investment, consumption, time schedules, random character of the audit etc.). 
Also, in multi-agent-based models, the entire mechanism of taxation, audit and fines of certain real or fictional countries 
can be simulated taking into account their specific legislation, organization, forms of payment and annual deadlines \cite{8}.
Our approach is less sophisticated concerning the financial aspects, 
but
its analytical character makes it more generally applicable than multi-agent simulations.
To the best of our knowledge, the tools and informations provided by our approach are 
neither gained 
nor obtainable
with
the methods of the other complexity economics papers which deal with tax evasion. First of all, 
these papers typically focus on the changes, induced by imitation or 
by some audit procedure, of agent moral attitude and behavior 
(see e.g. \cite{9,10,11,12,13},
and \cite{14} and \cite{15}, where also a review and a comparison of other references can be found).
Furthermore, when the effect of some audit is taken into account, no deterrent effect 
substantiated by an economic penalty is considered.
More in details, the models in \cite{9,10,11,12,13} constitute a variant of the Ising model 
of ferromagnetism: each citizen is represented by a spin variable,
which can be either in the tax compliant state $+1$ or in the tax evader state $-1$
(in \cite{12} a third state for undecided people is considered as well).
Citizens undergo transitions from $+1$ to $-1$, due to imitation of their nearest neighbours
or from $-1$ to $+1$, due to tax audits. Indeed, an audit on an evader is assumed to 
have as a consequence
the fact that she/he will remain honest for a certain number of steps.
While helpful for the analysis of evasion phenomena in relation to local interaction
and external controls, such an approach
does not allow conclusions on the effect of evasion on the income distribution as does our model.

\smallskip

In the remainder of the paper we start in Section \ref{model structure} with the proposal of a model structure 
capable of including the effects of some audit procedure; in practical terms we construct a system of ordinary differential equations
governing the time evolution of a distribution function over the income variable.
Then, in Section \ref{numerical solutions} we describe relevant features of numerical solutions 
of these equations.
Finally, in Section \ref{conclusions perspectives}
we carry out a critical analysis, we draw conclusions and mention as well 
further improvements of the model to be possibly developed in future work.

\section{A model structure accounting for audit procedure}
\label{model structure}

\subsection{The base structure for building a suitable model}

A structure
corresponding to a system of nonlinear ordinary differential equations
has been introduced in \cite{3} to model
taxation and redistribution in a society of individuals with different tax evasion behavior.
This structure
does not account for audit procedures, but we recall it here, because it represents the starting point for the construction of
the model discussed in this paper. It is expressed by
\begin{equation}
{{d x_j^{\alpha}} \over {d t}} =  
\sum_{h,k=1}^{n} \sum_{\beta,\gamma=1}^{m} {\Big (} C_{{(h,\beta)};{(k,\gamma)}}^{(j,\alpha)} 
+ T_{[{(h,\beta)};{(k,\gamma)}]}^{(j,\alpha)}(x) {\Big )}
x_h^{\beta} x_k^{\gamma}     -    x_j^{\alpha}  \sum_{k=1}^{n}  \sum_{\gamma=1}^{m} x_k^{\gamma} \, , 
\label{diffequationsJEIC}
\end{equation}
with $j = 1,...,n$ and $\alpha = 1,..,m$.
The equations in (\ref{diffequationsJEIC}) refer to a society of individuals
divided 
into a number $n$ of income {\it {classes}}
and 
into a number $m$ of {\it {sectors}} characterized by different evasion behaviors. Specifically,

$\bullet$ by $x_j^{\alpha}$ the fraction of $(j,\alpha)$-individuals is meant, namely the fraction of individuals who belong
to the $j$-th income class and to the $\alpha$-th evasion behavior sector.
The number of different individual {\it {groups}} is hence $n \times m$
and we denote $x$ the vector $x = (x_1^1,...,x_n^m) \in {\bf R}^{n\times m}$. 
The $j$-th income class is characterized by its average income $r_j$,
where $0 < r_1< r_2 < ... <\ r_n$. 
The total number of individuals as well as the tax evasion behavior of each individual
is supposed to remain constant in time.
In contrast, individuals may move through different income classes.

The definition and interpretation of the terms appearing on the r.h.s. of (\ref{diffequationsJEIC})
and describing the variation in time of $x_j^{\alpha}$ is as follows:

$\bullet$ for any $h, k, j = 1,...,n$ and any $\alpha, \beta, \gamma = 1,...,m$, the coefficient 
\begin{equation}
C_{{(h,\beta)};{(k,\gamma)}}^{(j,\alpha)} \in [0,+\infty)
\label{parameters C}
\end{equation}
expresses the probability that an $(h,\beta)$-individual will belong to 
the group $(j,\alpha)$ as a consequence of a direct interaction with an $(k,\gamma)$-individual.
These coefficients satisfy $\sum_{j=1}^{n} \sum_{\alpha=1}^{m} C_{{(h,\beta)};{(k,\gamma)}}^{(j,\alpha)} = 1$ 
for any fixed $(h,\beta)$, $(k,\gamma)$;

$\bullet$ for any $h, k, j = 1,...,n$ and any $\alpha, \beta, \gamma = 1,...,m$, the function
\begin{equation}
T_{[{(h,\beta)};{(k,\gamma)}]}^{(j,\alpha)}: {\bf R}^{n\times m} \to {\bf R}
\label{parameters T}
\end{equation}
expresses the variation in the group $(j,\alpha)$ 
due to the taxation and redistribution process associated to an interaction between an $(h,\beta)$-individual 
with an $(k,\gamma)$-individual. 
These functions are continuous and 
satisfy $\sum_{j=1}^{n} \sum_{\alpha=1}^{m}  T_{[{(h,\beta)};{(k,\gamma)}]}^{(j,\alpha)}(x) = 0$ 
for any fixed $(h,\beta)$, $(k,\gamma)$ and $x \in {\bf R}^{n\times m}$.

\smallskip

According to (\ref{diffequationsJEIC}), the variation in time of the individual number $x_j^{\alpha}$ is 
due to a whole of 
``microscopic"  interactions:
these include direct interactions of individuals who exchange money in pairs and 
indirect interactions 
which involve triplets of individuals and are
related to payment of taxes and redistribution of the tax revenue.
(see \cite{4} or \cite{5} for a detailed illustration of the mechanism).

\smallskip

Also a specific choice for the $C_{{(h,\beta)};{(k,\gamma)}}^{(j,\alpha)}$'s and the $T_{[{(h,\beta)};{(k,\gamma)}]}^{(j,\alpha)}$'s 
has been proposed in \cite{3}. In the following lines we recall it briefly.
To start with,

- let $p_{h,k}$ for $h, k = 1, ... , n$ be defined as
\begin{equation}
p_{h,k} = \min \{r_h,r_k\}/{4 r_n} \, ,
\label{phk}
\end{equation}
with the exception of the terms
$p_{j,j} = {r_j}/{2 r_n}$ for $j = 2, ..., n-1$,
$p_{h,1} = {r_1}/{2 r_n}$ for $h = 2, ..., n$, 
$p_{n,k} = {r_k}/{2 r_n}$ for $k = 1, ..., n-1$,
$p_{1,k} = 0$ for $k = 1, ..., n$
and
$p_{hn} = 0$ for $h = 1, ..., n$.
The coefficients (\ref{phk}) represent the probability that  in an encounter between an individual of the $h$-th  income class and one of the $k$-th class,
the one who pays is the former, see also \cite{5};

- let $S$, with $S << (r_{i+1} - r_{i})$ 
for all $i = 1, ..., n$, denote
an amount of money, namely the one which in each direct transaction one individual is supposed to pay to another;

- also, let $\tau_k$ for $k = 1, ... , n$ and with
$\tau_1 \le \tau_2 \le ... \le \tau_n$
denote the tax rate of the $k$-th income class
and let $\theta_{k,\alpha}$ for $k = 1, ..., n$ and $\alpha = 1, ..., m$
denote the effective tax rate\footnote{\ In the kind of tax evasion considered in this paper
the individual who receives money
reports a minor amount than she should.
In such a case, there is no advantage for the individual who pays.}, 
possibly reduced due to tax evasion, to be associated to any $(k,\alpha)$-individual; we assume that
\begin{equation}
\theta_{k,\alpha} = \theta_{ev}(\alpha) \, \tau_k \ , \nonumber
\label{ev tax rate}
\end{equation}
where $\theta_{ev}(\alpha)$ is a parameter in $[0,1] $. Of course, $\theta_{ev}(\alpha) = 0$ corresponds to total evasion and 
$\theta_{ev}(\alpha) = 1$ to absence of evasion.

Before giving the expressions of the $C_{{(h,\beta)};{(k,\gamma)}}^{(j,\alpha)}$'s and the $T_{[{(h,\beta)};{(k,\gamma)}]}^{(j,\alpha)}$'s,
we point out the following fact.

\medskip
\noindent {\it {Remark 1}}.
In a tax compliance case a direct interaction of
an individual of the $h$-th income class paying $S$ to one of the $k$-th class,
with this one paying the due tax, $S \, \tau_k$, to the government 
would be equivalent to that of the first individual paying an amount $S\, (1 - \tau_k)$ to the $k$-individual
and paying as well a quantity $S \, \tau_k$ to the government.
In a tax evasion case a direct interaction between a $(h,\beta)$-individual paying $S$ to a $(k,\alpha)$-individual
is equivalent to that of the $(h,\beta)$-individual paying an amount $S\, (1 - \theta_{k,\alpha})$ to the $(k,\alpha)$-individual
and paying as well a quantity $S \, \theta_{k,\alpha}$ to the government.
\medskip

\begin{figure*}
  \begin{center}
 \includegraphics[width=3.5cm,height=2.5cm]  {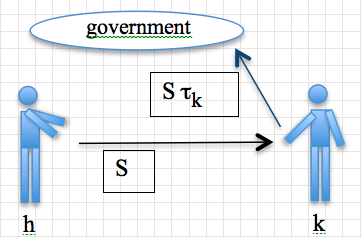}
  \hskip0.5cm
\includegraphics[width=3.5cm,height=2.5cm]  {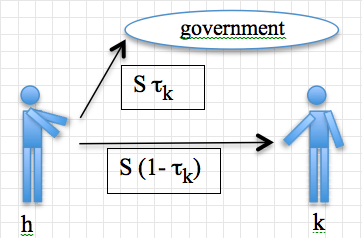}
  \hskip0.5cm
\includegraphics[width=3.5cm,height=2.5cm]  {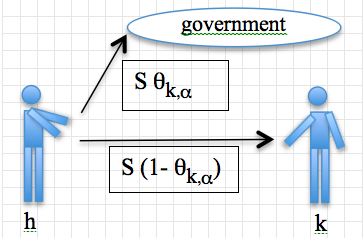}
  \end{center}
\caption{A sketchy illustration of some direct interactions. 
Left panel (absence of tax evasion): an $h$-individual pays $S$ to a $k$-one,
and this one pays the due tax, $S \, \tau_k$, to the government.
Central panel (equivalent situation):
the $h$-individual pays $S\, (1 - \tau_k)$ to the $k$-one
and pays as well a quantity $S \, \tau_k$ to the government. 
Right panel (situation equivalent to one with tax evasion): the $h$-individual pays $S\, (1 - \theta_k)$ to a $(k,\alpha)$-individual, 
and 
pays as well a quantity $S \, \theta_k$ to the government.} 
\end{figure*}

\smallskip

The parameters (\ref{parameters C}) and (\ref{parameters T}) proposed in \cite{3} are as follows. 

\smallskip

Among the coefficients $C_{{(h,\beta)};{(k,\gamma)}}^{(j,\alpha)}$, the only nonzero ones are:
\begin{eqnarray}
\label{definitionofC}
C_{{(j+1,\alpha)};{(k,\beta)}}^{(j,\alpha)} & = 
                  & p_{j+1,k} \, \frac{S \, (1-\theta_{k,\beta}) }{r_{j+1} - r_{j}} \, , \\ \nonumber
C_{{(j,\alpha)};{(k,\beta)}}^{(j,\alpha)} & = 
            & 1 
               - \, p_{k,j} \, \frac{S \, (1-\theta_{j,\alpha})}{r_{j+1} - r_{j}} 
               - \, p_{j,k} \, \frac{S \, (1-\theta_{k,\beta})}{r_{j} - r_{j-1}} \, ,  \\           
C_{{(j-1,\alpha)};{(k,\beta)}}^{(j,\alpha)} & = 
               & p_{k,j-1} \, \frac{S \, (1-\theta_{j-1,\alpha})}{r_{j} - r_{j-1}} \, , \nonumber
\end{eqnarray} 
where the indices $\alpha$ and $\beta$ take any value in $1, ..., m$ and, 
as for the indices $j$ and $k$, 
each term is defined only when it is meaningful: e.g., 
the second addendum on the r.h.s. in the second line of (\ref{definitionofC}) 
is present only for $j \le n-1$ and $k \ge 2$, whereas the third addendum is present only for $j \ge 2$ and $k \le n-1$.

\smallskip

The functions $T_{[{(h,\beta)};{(k,\gamma)}]}^{(j,\alpha)}$ take the form
\begin{eqnarray}
\label{definitionofT}
T_{[{(h,\beta)};{(k,\gamma)}]}^{(j,\alpha)}(x) 
& = & 
\frac{p_{h,k} \, S \, \theta_{k,\gamma}} {\sum_{i=1}^{n} \sum_{\lambda=1}^{m} x_{i}^{\lambda}} {\bigg (}  \frac{x_{j-1}^{\alpha}}{(r_j - r_{j-1})} 
-   
\frac{x_{j}^{\alpha}}{(r_{j+1} - r_{j})} {\bigg )} \\
\ & + &  p_{h,k} \, S \, \theta_{k,\gamma} \, 
{\bigg (} 
\frac{\delta_{h,j+1}{\delta_{\alpha,\beta}}}{r_h - r_{j}} \, - \, \frac{\delta_{h,j}{\delta_{\alpha,\beta}}}{r_h - r_{j-1}}
{\bigg )} 
\, \frac{{\sum_{i=1}^{n-1} \sum_{\lambda=1}^{m} x_{i}^{\lambda}}}{{\sum_{i=1}^{n} \sum_{\lambda=1}^{m} x_{i}^{\lambda}}} \, ,  \nonumber 
\end{eqnarray} 
with $\delta_{i,j}$ denoting the {\it Kronecker delta} and, as in (\ref{definitionofC}), with the understanding that
each term is defined only when meaningful.
We recall that,
for technical reasons,
the effective amount of money 
paid as tax is multiplied by 
$({\sum_{i=1}^{n-1} \sum_{\lambda=1}^{m} x_{i}^{\lambda}})/({\sum_{i=1}^{n} \sum_{\lambda=1}^{m} x_{i}^{\lambda}})$.

\medskip
\noindent {\it {Remark 2}}.
As observed in \cite{3}, conservation in time of both the number of individuals and the global income hold true
for (\ref{diffequationsJEIC}). Equivalently,

$\bullet$
For any initial condition $x_0 = \{{x_{0_j}^{\alpha}}\}_{j=1,...n;\alpha=1,...m}$
having non negative components $x_{0_j}^{\alpha}$, such that 
${\sum_{j=1}^{n} \sum_{\alpha=1}^{m} x_{0_j}^{\alpha} = 1}$\footnote{\ The number $1$ is chosen here as a normalization value.},
a unique solution $x(t) =  \{{x_{0_j}^{\alpha}}(t)\}_{j=1,...n;\alpha=1,...m}$ of $(\ref{diffequationsJEIC})$ exists,
which is defined for all $t \in [0,+\infty)$, satisfies $x(0) = x_0$ and also
\begin{equation}
x_{j}^{\alpha}(t) \ge 0 \ \hbox{for} \ j = 1, ... , n \ \hbox{and} \ \alpha = 1, ... , m 
\ \hbox{and} \
{\sum_{j=1}^{n} \sum_{\alpha=1}^{m} x_{j}^{\alpha}(t) = 1} \ \hbox{for all} \ t \ge 0 \, . 
\label{buonaposizione}
\end{equation}
This implies that the right hand sides of system $(\ref{diffequationsJEIC})$ simplify;
they are in fact polynomials of third degree.

$\bullet$ 
The scalar function
$\mu(x)=\sum_{j=1}^n r_j \sum_{\alpha=1}^{m} x_j^{\alpha}$
remains constant along each solution of system $(\ref{diffequationsJEIC})$.

\subsection{Incorporating audit procedure terms into the model}

We are now ready to incorporate also the effects of some audit procedure into the model.
Towards this aim
we will introduce here two further parameters, by means of which
the equations (\ref{diffequationsJEIC}) will be suitably generalized. 
These parameters will be denoted by $\sigma \in [0,1]$ and $\xi >1$. They
represent respectively 
\begin{itemize}
\item the fraction $\sigma$ of individuals in the society who are subject to an audit,
\item a multiplicative factor $\xi$ expressing the coefficient by which the amount of 
tax evaded has to be multiplied to give the wanted fine.
\end{itemize}

To explain the generalized equation form we argue as follows.
First of all, it is convenient to split the r.h.s. of the equations into two parts,
which represent respectively the contribution,
both direct and indirect, of the non-audited and of the audited individuals. In the first part,
all terms describing the binary and the ternary interactions are exactly
as in the case without audit.
The second part maintains the structure of the first one, but some coefficients in it are different.
Arguing similarly as in Remark $1$, we observe here that in the presence of an audit procedure
in a case with possible tax evasion, 
a direct interaction between a $(h,\beta)$-individual paying $S$ to a $(k,\alpha)$-individual
can be thought of as equivalent to that of the $(h,\beta)$-individual paying an amount 
$S\, \Big(1 - \big(\theta_{k,\alpha} + \xi (\tau_k - \theta_{k,\alpha})\big)\Big)$ to the $(k,\alpha)$-individual
and paying as well a quantity 
$S \, \big(\theta_{k,\alpha} + \xi (\tau_k - \theta_{k,\alpha})\big)$ to the government.
Indeed, the final effect of the operation is that

- the $(h,\beta)$-individual pays in total the amount $S$,

\noindent and

- the amount corresponding to the tax revenue related to this interaction is $S \, \big(\theta_{k,\alpha} + \xi (\tau_k - \theta_{k,\alpha})\big)$.
If the $(k,\alpha)$-individual is tax compliant, i.e. $\theta_{k,\alpha} = \tau_k$, this amount coincides with $S \, \tau_k$;
otherwise, due to the fact that $\xi > 1$,
this amount is greater than $S \, \tau_k$. (It would have been smaller than $S \, \tau_k$ 
if evasion without audit had occurred).

\smallskip

We denote from now on by
$$
{C_{\xi}}_{{(h,\beta)};{(k,\gamma)}}^{(j,\alpha)}
\qquad \hbox{and} \qquad 
{T_{\xi}}_{[{(h,\beta)};{(k,\gamma)}]}^{(j,\alpha)}(x)
$$
the coefficients and the functions defined as in
(\ref{definitionofC}) and (\ref{definitionofT}), with the difference that any $\theta_{k,\alpha}$
appearing
in (\ref{definitionofC}) and (\ref{definitionofT}) must be replaced by
$\theta_{k,\alpha} + \xi (\tau_k - \theta_{k,\alpha})$.
Similarly to the $C_{{(h,\beta)};{(k,\gamma)}}^{(j,\alpha)}$'s and $T_{[{(h,\beta)};{(k,\gamma)}]}^{(j,\alpha)}$'s
in (\ref{definitionofC}) and (\ref{definitionofT}),
the coefficients ${C_{\xi}}_{{(h,\beta)};{(k,\gamma)}}^{(j,\alpha)}$ and the functions ${T_{\xi}}_{[{(h,\beta)};{(k,\gamma)}]}^{(j,\alpha)}$
satisfy
$$
\sum_{j=1}^{n} \sum_{\alpha=1}^{m} {C_{\xi}}_{{(h,\beta)};{(k,\gamma)}}^{(j,\alpha)} = 1
\qquad \hbox{and} \qquad 
\sum_{j=1}^{n} \sum_{\alpha=1}^{m}  {T_{\xi}}_{[{(h,\beta)};{(k,\gamma)}]}^{(j,\alpha)}(x) = 0 \ ,
$$
for any fixed $(h,\beta)$, $(k,\gamma)$ and $x \in {\bf R}^{n\times m}$.

The equations of the audit model can now be written. They take the form 
\begin{eqnarray}
\label{diffequationsAudit}
{{d x_j^{\alpha}} \over {d t}} 
& = &   
\big( 1 - \sigma \big) \ 
\bigg(
\sum_{h,k=1}^{n} \sum_{\beta,\gamma=1}^{m} {\Big (} C_{{(h,\beta)};{(k,\gamma)}}^{(j,\alpha)} 
+ T_{[{(h,\beta)};{(k,\gamma)}]}^{(j,\alpha)}(x) {\Big )}
x_h^{\beta} x_k^{\gamma}     -    x_j^{\alpha} 
\bigg) \\ \nonumber 
\ & + &
\sigma \ 
\bigg(
\sum_{h,k=1}^{n} \sum_{\beta,\gamma=1}^{m} {\Big (} {C_{\xi}}_{{(h,\beta)};{(k,\gamma)}}^{(j,\alpha)} 
+ {T_{\xi}}_{[{(h,\beta)};{(k,\gamma)}]}^{(j,\alpha)}(x) {\Big )}
x_h^{\beta} x_k^{\gamma}     -    x_j^{\alpha} 
\bigg)
\, , \nonumber 
\end{eqnarray} 
with $j = 1,...,n$ and $\alpha = 1,..,m$, with the terms on the right hand side reducing, in view of the Remark $2$, 
to polynomials of third degree.

\medskip

\noindent {\it {Remark 3}}.
The penalty inflicted in this model refers to a single economic transaction:
no reference is done to the accumulated capital for it to be paid.
Accordingly, in order to guarantee non negativity of the quantity
$S\, \Big(1 - \big(\theta_{k,\alpha} + \xi (\tau_k - \theta_{k,\alpha})\big)\Big)$ 
which a $(h,\beta)$-individual pays in a generic transaction to a $(k,\alpha)$-individual,
we have to impose some constraint on the parameter $\xi$.
We will assume from now on that
\begin{equation}
\tau_n \le 50 \% \qquad \hbox{and} \quad \xi \in (1,2] \, .
\label{boundontaumaxandxi}
\end{equation}

\section{Numerical solutions and results}
\label{numerical solutions}

It is a straightforward exercise to check that
the two properties stated in Remark $2$, i.e. the existence for a given initial condition of a unique solution 
satisfying $(\ref{buonaposizione})$ and the conservation in time of the global income,
hold true for the system $(\ref{diffequationsAudit})$ as well.
In addition, the running of several numerical solutions provides evidence of the following fact.

\medskip

\noindent {\it {Property 1}}. 
If the parameters of the model (namely, $r_1, ..., r_n$, $S$, the $\tau_k$'s, the $\theta_{ev}(\alpha)$'s),
the fraction of individuals 
with different behavior\footnote{\ The fraction of individuals with a specific evasion behavior is assumed here to be the same in each income class.} 
and
$\mu \in [r_1,r_n]$ are all fixed, then the solutions $x(t) =  \{{x_{0_j}^{\alpha}}(t)\}_{j=1,...n;\alpha=1,...m}$ evolving from initial conditions 
$x_0 = \{{x_{0_j}^{\alpha}}\}_{j=1,...n;\alpha=1,...m}$, 
for which $x_{0_j}^{\alpha} \ge 0$ for any $j = 1, ... , n$ and $\alpha = 1, ... , m$, 
$$
{\sum_{j=1}^{n} \sum_{\alpha=1}^{m} x_{0_j}^{\alpha} = 1}
\qquad \hbox{and} \qquad
\sum_{j=1}^n  \sum_{\alpha=1}^{m} r_j \, x_{0_j}^{\alpha} = \mu \, ,
$$
tend asymptotically to a same stationary distribution
as $t \to +\infty$.

\medskip

We are specifically interested into the effects of different audit policies on the lung run income distribution. 
Hence, we compare various scenarios emerging in correspondence to different 
percentages $\sigma$ of the audited individuals and different penalty coefficients $\xi$.

\medskip

To obtain numerical solutions, one has to fix the parameters of the model. We list next 
the specific choice we carried out in this connection and then we'll
give account of the output of a number of numerical solutions.

We took the number of income classes $n = 9$, the number of behavioral sectors $m=3$, the average incomes $r_j = 10 j$ for $j = 1, ... n$, 
the unit amount of exchanged money $S = 0.1$, and the tax rates
\begin{equation}
\tau_j = \tau_{1} +  \frac{j - 1}{n-1} \, (\tau_{n} - \tau_{1}) \, , \qquad \hbox{for} \ j = 1, ... , n \, 
\label{progressivetaxrates}
\end{equation}
with $\tau_1 = 23 \%$ and $\tau_n = 43 \%$\footnote{\ These are the values of the minimal and of the maximal tax rate relative to IRPEF in Italy.}.
Also, we considered the case in which each one of the three 
evasion-type behavioral sectors contains exactly one third of the population. We report below the 
results obtained in correspondence to 

\smallskip

(i) cases in which
\begin{itemize}
\item the first sector consists of compliant individuals,
\item the second sector consists of individuals who use to evade half of the taxes they should,
\item the third sector consisting of individuals who use to evade three quarters of the taxes they should,
\end{itemize}
corresponding to the choice of the parameters
\begin{equation}
\theta_{ev}(1) = 1 \, , \qquad \theta_{ev}(2) = 1/2 \, , \qquad \theta_{ev}(3) = 1/4 \, ,
\label{example of theta evasion105025}
\end{equation}

and 

\smallskip

(ii) cases in which there is less evasion and specifically
the evasion rate is half the evasion rate of the previous cases, namely cases in which
\begin{itemize}
\item the first sector consists of compliant individuals,
\item the second sector consists of individuals who use to evade one quarter of the taxes they should,
\item the third sector consisting of individuals who use to evade three eights of he taxes they should,
\end{itemize}
corresponding to the choice of the parameters
\begin{equation}
\theta_{ev}(1) = 1 \, , \qquad \theta_{ev}(2) = 3/4 \, , \qquad \theta_{ev}(3) = 5/8 \, .
\label{example of theta evasion10750625}
\end{equation}

For each of the two situations
described by the parameter choices $(\ref{example of theta evasion105025})$ and $(\ref{example of theta evasion10750625})$
we considered a range of different cases corresponding to different values of the parameters $\sigma$ and $\xi$.
In fact, we ran several sets of simulations, with each 
set characterized by the value $\mu$ of the constant global income.
We evaluated the Gini index and the tax revenue relative to the 
stationary distribution reached in the long run by the solutions of the systems $(\ref{diffequationsAudit})$ for different values of 
$\sigma \in [0,1]$ and $\xi \in (1,2]$.
We recall here that 

\begin{itemize}
\item the {\it{Gini index}} provides a measure of the extent to which a distribution - here, the income distribution -
deviates from a perfectly equal one. It is graphically represented by the area between the Lorenz curve 
(which plots the cumulative income share of a population on the $y$ axis
against the distribution of the population on the $x$ axis)
and the line of perfect equality.
It takes values in $[0,1]$, where $0$ represents the complete equality and $1$ the maximal inequality;
\item the {\it{tax revenue}} is the income collected by the government through taxation.
Here, we calculate in fact the tax revenue in the unit time as 
\begin{equation}
{\rm{T R}} = \sum_{h=1}^{n} \, \sum_{k=1}^{n} \, \sum_{j=1}^{n - 1} \, \sum_{\alpha=1}^{m} \, \sum_{\beta=1}^{m} \, \sum_{\gamma=1}^{m} \, 
\, S \, p_{hk} \,\Big(\theta_{k,\gamma} + \sigma \, \xi \, (\tau_k - \theta_{k,\gamma})\Big) \, 
x_{j}^{\alpha} \, x_{h}^{\beta} \, x_{k}^{\gamma} \, , \nonumber
\end{equation}
where ${x}_i^{\lambda}$ denotes the fraction of individuals in the $i$-th income class, $\lambda$-th sector at equilibrium. 
\end{itemize}

The results of the simulations are all qualitatively consistent with those summarized in Tables $\ref{table1}$ and $\ref{table2}$. 
The variations of the values of ${\rm{T R}}$ given there are very small, but, one should not forget that they just provide
the tax revenue in the unit time. The tax revenue of a definite period coincides with a sum
of contributions of this kind.
Looking at the numbers contained in Tables $\ref{table1}$ and  $\ref{table2}$, one may notice at a first glance that

\smallskip

\noindent $\bullet$ for fixed $\xi$, the tax revenue increases and the Gini index decreases as $\sigma$ increases,

\noindent $\bullet$ for fixed $\sigma$, the tax revenue increases and the Gini index decreases as $\xi$ increases.

\smallskip

\begin{table}[ht]
\begin{center}
{The data in this table refer to the asymptotic stationary income distributions 
of systems characterized by different values of the 
fraction $\sigma$ of audited individuals and of the penalty multiplicative factor $\xi$, 
in cases in which the total income is $\mu = 79$
and
$\theta_{ev}(1) = 1, \theta_{ev}(2) = 1/2, \theta_{ev}(3) = 1/4$.
The value of the Gini index and of the tax revenue is provided in each case. Their value in the case with no evasion
would be respectively: 
{\rm{GI}} = $0.367068$ and {\rm{TR}} = $1.789 \cdot 10^{-3}$,
and in the case with evasion, but no audit 
{\rm{GI}} = $0.383093$ and {\rm{TR}} = $0.949 \cdot 10^{-3}$.}
{\begin{tabular}{@{}l|ccccc@{}} 
\toprule
{\bf{Gini Index}} \ & $\xi = 1.25$ \ &\ $\xi = 1.40$ \ & \ $\xi = 1.55$ \ & \ $\xi = 1.70$ \ &\ $\xi = 1.85$ \nonumber \\
\hline
\quad \\
\quad \ $\sigma = {2}/{56}$ \ & \ 0.382193  \ & \ 0.382086   \ & \ 0.381979  \ & \ 0.381873   \ & \ 0.381766   \nonumber \\
\quad \ $\sigma = {5}/{56}$ \ & \ 0.380873  \ & \ 0.380614  \ & \ 0.380355 \ & \ 0.380099   \ & \ 0.379844   \nonumber \\
\quad \ $\sigma = {8}/{56}$ \ & \ 0.37959  \ & \ 0.379187  \ & \ 0.378789 \ & \ 0.378394  \ & \ 0.378003  \nonumber \\
\quad \ $\sigma = {11}/{56}$ \ & \ 0.378345  \ & \ 0.377809  \ & \ 0.377281 \ & \ 0.37676   \ & \ 0.376247   \nonumber \\
\quad \ $\sigma = {14}/{56}$ \ & \ 0.377138  \ & \ 0.376479  \ & \ 0.375833 \ & \ 0.375198  \ & \ 0.374577  \nonumber \\
\  \ & \ \  \ & \  \   \ & \ \   \nonumber \\
\hline
{\bf{Tax Revenue}} \ & \ $\xi = 1.25$ \ &\ $\xi = 1.40$ \ & \ $\xi = 1.55$ \ & \ $\xi = 1.70$ \ &\ $\xi = 1.85$ \nonumber \\
\hline
\quad \\
\quad \ $\sigma = {2}/{56}$ \ & \ $0.988 \cdot 10^{-3}$ \ & \ $0.992 \cdot 10^{-3}$  \ & \ $0.997 \cdot 10^{-3}$ \ & \ $1.002 \cdot 10^{-3}$  \ & \ $1.006 \cdot 10^{-3}$  \nonumber \\
\quad \ $\sigma = {5}/{56}$ \ & \ $1.045 \cdot 10^{-3}$  \ & \ $1.057 \cdot 10^{-3}$  \ & \ $1.068 \cdot 10^{-3}$ \ & \ $1.080 \cdot 10^{-3}$  \ & \ $1.091 \cdot 10^{-3}$  \nonumber \\
\quad \ $\sigma = {8}/{56}$ \ & \ $1.103 \cdot 10^{-3}$  \ & \ $1.121 \cdot 10^{-3}$  \ & \ $1.139 \cdot 10^{-3}$ \ & \ $1.158 \cdot 10^{-3}$  \ & \ $1.176 \cdot 10^{-3}$  \nonumber \\
\quad \ $\sigma = {11}/{56}$ \ & \ $1.160 \cdot 10^{-3}$  \ & \ $1.185 \cdot 10^{-3}$  \ & \ $1.210 \cdot 10^{-3}$ \ & \ $1.235 \cdot 10^{-3}$  \ & \ $1.260 \cdot 10^{-3}$  \nonumber \\
\quad \ $\sigma = {14}/{56}$ \ & \ $1.217 \cdot 10^{-3}$  \ & \ $1.249 \cdot 10^{-3}$  \ & \ $1.281 \cdot 10^{-3}$ \ & \ $1.313 \cdot 10^{-3}$  \ & \ $1.344 \cdot 10^{-3}$  \nonumber \\
\botrule
\end{tabular} \label{table1}}
\end{center}
\end{table}

\begin{table}[ht]
\begin{center}
{The data in this table refer to the asymptotic stationary income distributions 
of systems characterized by
different values of the 
fraction $\sigma$ of audited individuals and of the penalty multiplicative factor $\xi$, 
in cases in which the total income is $\mu = 79$
and
$\theta_{ev}(1) = 1, \theta_{ev}(2) = 3/4, \theta_{ev}(3) = 5/8$.
The value of the Gini index and of the tax revenue is provided in each case. Their value in the case with no evasion
would be respectively: 
{\rm{GI}} = $0.367068$ and {\rm{TR}} = $1.789 \cdot 10^{-3}$,
and in the case with evasion, but no audit 
{\rm{GI}} = $0.373967$ and {\rm{TR}} = $1.376 \cdot 10^{-3}$.}
{\begin{tabular}{@{}l|ccccc@{}} 
\toprule 
{\bf{Gini Index}} \ & \ $\xi = 1.25$ \ &\ $\xi = 1.40$ \ & \ $\xi = 1.55$ \ & \ $\xi = 1.70$ \ &\ $\xi = 1.85$ \nonumber \\
\hline
\quad \\
\quad \ $\sigma = {2}/{56}$ \ & \ 0.373611  \ & \ 0.373568  \ & \ 0.373526 \ & \ 0.373483  \ & \ 0.373441  \nonumber \\
\quad \ $\sigma = {5}/{56}$ \ & \ 0.373084  \ & \ 0.37298  \ & \ 0.372876 \ & \ 0.372773  \ & \ 0.37267  \nonumber \\
\quad \ $\sigma = {8}/{56}$ \ & \ 0.372567  \ & \ 0.372404  \ & \ 0.372242 \ & \ 0.372081  \ & \ 0.371921  \nonumber \\
\quad \ $\sigma = {11}/{56}$ \ & \ 0.372061  \ & \ 0.371841  \ & \ 0.371624 \ & \ 0.371408  \ & \ 0.371194  \nonumber \\
\quad \ $\sigma = {14}/{56}$ \ & \ 0.371565  \ & \ 0.371291  \ & \ 0.371021 \ & \ 0.370754  \ & \ 0.37049  \nonumber \\
\  \ & \ \  \ & \  \   \ & \ \   \nonumber \\
\hline
{\bf{Tax Revenue}} \ & \ $\xi = 1.25$ \ &\ $\xi = 1.40$ \ & \ $\xi = 1.55$ \ & \ $\xi = 1.70$ \ &\ $\xi = 1.85$ \nonumber \\
\hline
\quad \\
\quad \ $\sigma = {2}/{56}$ \ & \ $1.395 \cdot 10^{-3}$ \ & \ $1.397 \cdot 10^{-3}$  \ & \ $1.399 \cdot 10^{-3}$ \ & \ $1.401. \cdot 10^{-3}$  \ & \ $1.404 \cdot 10^{-3}$  \nonumber \\
\quad \ $\sigma = {5}/{56}$ \ & \ $1.423 \cdot 10^{-3}$  \ & \ $1.428 \cdot 10^{-3}$  \ & \ $1.434 \cdot 10^{-3}$ \ & \ $1.440 \cdot 10^{-3}$  \ & \ $1.445 \cdot 10^{-3}$  \nonumber \\
\quad \ $\sigma = {8}/{56}$ \ & \ $1.451 \cdot 10^{-3}$  \ & \ $1.460 \cdot 10^{-3}$  \ & \ $1.469 \cdot 10^{-3}$ \ & \ $1.478 \cdot 10^{-3}$  \ & \ $1.487 \cdot 10^{-3}$  \nonumber \\
\quad \ $\sigma = {11}/{56}$ \ & \ $1.479 \cdot 10^{-3}$  \ & \ $1.491 \cdot 10^{-3}$  \ & \ $1.503 \cdot 10^{-3}$ \ & \ $1.516 \cdot 10^{-3}$  \ & \ $1.528 \cdot 10^{-3}$  \nonumber \\
\quad \ $\sigma = {14}/{56}$ \ & \ $1.507 \cdot 10^{-3}$  \ & \ $1.522 \cdot 10^{-3}$  \ & \ $1.538 \cdot 10^{-3}$ \ & \ $1.553 \cdot 10^{-3}$  \ & \ $1.569 \cdot 10^{-3}$  \nonumber \\
\botrule
\end{tabular} \label{table2}}
\end{center}
\end{table}

If such a behavior is definitely reasonable form a qualitative point of view, the added value 
of the model is given by the quantitative aspect.
One can for example estimate the amount of money collected
in correspondence to different choices of the parameters $\xi$ and $\sigma$.
And, as illustrated in the next lines, also values of these parameters different from those 
appearing in the tables above can be taken into account.
Indeed, looking more carefully at the Tables $\ref{table1}$ and $\ref{table2}$,
one realizes that both the Gini index and the tax revenue values depend linearly on $\xi$ and
depend linearly on $\sigma$.
In other words, one finds that both the Gini index and the tax revenue may be approximated by bilinear functions of the
variables $\xi$ and $\sigma$. 
One can then employ a {\it{least squares}} method to find a surface coinciding with the graph of a bilinear function\footnote{\ That a bilinear function 
as in (\ref{bilinear function}) already provides a good approximation, it can be checked
by  trying and looking
instead for a function
$f(\xi,\sigma) =  a_0 + a_{10} \, \xi + a_{01} \, \sigma + a_{11} \, \xi \, \sigma + a_{20} \, \xi^2 + a_{02} \, \sigma^2$. 
When this is done and a least squares procedure is performed,
the resulting coefficients of $\xi^2$ and $\sigma^2$ turn out to be of $1$ to $3$ magnitude order less than each other coefficient
in the expression of $f$.}
\begin{equation}
f(\xi,\sigma) =  a_0 + a_{10} \, \xi + a_{01} \, \sigma + a_{11} \, \xi \, \sigma 
\label{bilinear function}
\end{equation}
best fitting the discrete data available.
In this connection, denote
$$
\xi_1 = 1.25, \ \xi_2 = 1.40,  \  \xi_3 = 1.55, \ \xi_4 = 1.70, \ \xi_5 = 1.85 \, ,
$$
and
$$
\sigma_1 = 2/56, \ \sigma_2 = 5/56, \ \sigma_3 = 8/56, \ \sigma_4 = 11/56, \ \sigma_5 = 14/56 \, .
$$
Searching, for example with reference to Table $\ref{table1}$, 
values of the coefficients $a_0, a_{10}, a_{01}, a_{11}$ in (\ref{bilinear function}), to which a minimum of
\begin{equation}
\sum_{h=1}^5 \, \sum_{k=1}^5 \, \big(f(\xi_{h},\sigma_{k}) -  {\rm{TR}}(\xi_{h},\sigma_{k})\big)^2
\end{equation}
corresponds,
and constructing with these values the function in (\ref{bilinear function}), one finds for it
the expression 
$$
5.447  \cdot 10^{-4} + 3.521  \cdot 10^{-4} \, \xi - 9.710 \cdot 10^{-4} \, \sigma + 8.464 \cdot 10^{-4} \, \xi \, \sigma \, .
$$
In conjunction with an estimate of the auditing costs, this allows
to calibrate the parameters $\xi$ and $\sigma$ so as to have effective auditing and fine.
Indeed, by fixing a reasonable value of the expected tax revenue, say $C$, one may look for the 
coordinates $\xi$ and $\sigma$ which satisfy
\begin{equation}
5.447  \cdot 10^{-4} + 3.521  \cdot 10^{-4} \, \xi - 9.710 \cdot 10^{-4} \, \sigma + 8.464 \cdot 10^{-4} \, \xi \, \sigma = C \, .
\end{equation}
This equation defines a curve and,
as the next formulae show, each of the coordinates can be expressed as a function of the other one:
\begin{eqnarray}
\label{xiandsigmacurves}
\xi & = & \frac{C - 5.447  \cdot 10^{-4} + 9.710  \cdot 10^{-4}  \, \sigma}{3.521 \cdot 10^{-4}  + 8.464  \cdot 10^{-4}  \, \sigma} 
= \frac{\tilde C - 5.447 + 9.710  \, \sigma}{3.521 + 8.464  \, \sigma} 
 \, , \\ \nonumber
\sigma & = & \frac{C - 5.447  \cdot 10^{-4} - 3.521 \cdot 10^{-4} \, \xi}{- 9.710  \cdot 10^{-4} + 8.464  \cdot 10^{-4} \, \xi}
= \frac{\tilde C - 5.447  - 3.521 \, \xi}{- 9.710 + 8.464 \, \xi}
\, ,
\end{eqnarray}
where $\tilde C = C \cdot 10^{4}$.
The equations $(\ref{xiandsigmacurves})$ can 
give some insight as how to fix the parameters $\xi$ and $\sigma$ in an appropriately {\it{balanced}} way, i.e. 
in such a way that neither $\xi$ nor $\sigma$ are too large.

\section{Critical analysis and conclusions}
\label{conclusions perspectives}

This paper proposes a dynamic model for the description of processes of
taxation, redistribution and auditing in the presence of tax evasion, possibly occurring in different measure.
The focus is on the effects of different enforcement regimes on the income distribution.
According to the model here formulated, the main role to this end
is played by two parameters: one of them, $\sigma$, expresses
the fraction of audited taxpayers; the other one, $\xi$, is
the coefficient by which the amount of owed tax must be multiplied
to give the tax fraud penalty.
The model fits in with a complex system perspective: society is treated as a large collection of
heterogeneous individuals,
divided into income classes and characterized by different compliance and non-compliance behaviors. 
Total wealth is supposed to remain constant in time and the evolution of the income distribution is described by
a system of nonlinear ordinary differential equations which involve transition probabilities. 
The long-run solutions obtained with a number of numerical simulations show 
stationary aggregate-level income distributions emerging from the whole of individual interactions.

\smallskip

An aspect of interest of this model is the possibility of employing it as an ``explorative'' tool.
Simulations corresponding to different conceivable values of the mentioned parameters 
can provide insight as to whether
it is more convenient to increase the severity of punishment or the number of audits.
Indeed, they allow to forecast the emergence of different scenarios.  
Vice versa, the model can be also used to derive, in correspondence to a reasonable desired value $C$ of the tax revenue,
the relation between the parameters $\xi$ and $\sigma$ which are compatible with $C$. And this,
combined with the intent of keeping  under control the values of these parameters,
can suggest convenient choices of these parameters.

 \smallskip
 
The tax audits and fines described by our model can be considered as only moderately ``punitive''. We assume that tax evaders, 
if found guilty in an audit, must pay a fine which cuts their gain in the current transaction, while the rest of their wealth is not affected 
and their evasion habits also remain unchanged. This limitation of the model can be lifted in a more general version; actually, 
in the general formulation of the equations (\ref{diffequationsJEIC}) the elements $C^{(j,\alpha)}_{(h,\beta);(k,\gamma)}$
of the transition matrices are not restricted to be proportional to $\delta_{\alpha \beta}$ (namely, different from zero only if $\alpha = \beta$)
as is the case in the particular model investigated here.
On the other hand,
considering the amount of the fines, it is easy to check 
that the persistence of the individuals' evasion behavior is coherent with their subjective perception of economic advantage. 
In fact, even when the audits have maximum frequency ($\sigma=7/28$) and largest penalty rate ($\xi=185/100$), 
if also evaders know this in advance, they may conclude that it is still more convenient for them to evade and risk to pay the fine.

 \smallskip

From the point of view of the state administration the audits can produce a remarkable increase in the tax revenue. 
For instance, considering again the maximum audit frequency and maximum penalty, and assuming evasion levels equal to $0$, $1/2$ and $3/4$ 
in the three behavioral sectors as in Table $\ref{table1}$, the increase in the tax revenue with respect to the case of no audits is about $40\%$. 
As pointed out in Section $\ref{numerical solutions}$,
the dependence of this figure on the parameters $\sigma$ and $\xi$ is approximately linear, so a hypothetical policy-maker 
could use the model and some additional information on the costs of the audits in order to assess the most convenient audits/fines policy. 
We recall that the taxation level implemented in our model is quite realistic: comparing ``pre-redistribution'' values of $G$ 
(those obtained when taxation terms are switched off) with the values after redistribution \cite{16} it turns out that we are quite close 
to the real data of the United States (while, for instance, economies like Germany and Denmark exhibit a markedly larger redistribution gap).

\smallskip

A priori one could think that audits and fines should have a positive effect on the reduction of economic inequality 
and correspondingly of the Gini index $G$. According to our model, however, such effect is small: 
the decrease of $G$ when the audit frequency $\sigma$ and penalty ratio $\xi$ are increased is only of the order of $0.1\%$. 
In contrast, the decrease of $G$ following a diminution of the evasion level \cite{2} or a weighted redistribution of welfare 
in favor of the poor classes \cite{16} is of the order of $1\%$ or more. In view of the strong dependence 
of the tax revenue on $\sigma$ and $\xi$, this is a further confirmation that the relation between the Gini index 
and the tax revenue is not simple, and is influenced by several factors. 
In \cite{16}, for instance, we found that when the gap in the tax rates varies from $30\%$ - $45\%$ to $10\%$ - $65\%$,
the tax revenue decreases and so does the Gini index (arguably because, although the rich pay more and the poor less, 
the largest part of the tax revenue still comes from the poor and middle classes). 
On the other hand, when the welfare redistribution parameters are changed in favor of the poor classes, the tax revenue increases and $G$ decreases.
Besides, we would like to recall that in the presence of a high level of tax evasion even the sizeable increase of the Gini index 
does not adequately reflect the injustice actually present in the society; we found in \cite{3} for instance that with an evasion level of $40\%$ 
tax evaders can enjoy an average wealth gap up to ca.\ $30\%$ with respect to non-evaders.

\smallskip

Finally, we emphasize again that the model requires a condition:
in fact, its two main parameters cannot be totally arbitrary. They have to satisfy a constraint, which for simplicity
has been expressed here 
as the need of the inequalities $\tau_n < 50\%$ and $\xi \le 2$ to hold true, although more general formulations are possible.
It would be desirable to get rid of this condition.
Pointing to directions of possible future research, we also observe that a relevant step forward would be the
inclusion in the model of the possibility for individuals
of changing behavioral type in time for different reasons, among which e.g. auditing experience.






\begin{thebibliography}{00}

\bibitem{1} C. Schinckus, 
Between complexity of modelling and modelling of complexity: an essay of econophysics,
{\it {Physica A}} {\textbf 392}, 3654-3665 (2013).
\bibitem{2} M.L. Bertotti, G. Modanese, 
Micro to macro models for income distribution in the absence and in the presence of tax evasion,
{\it{Appl. Math. Comput.}} \textbf{244}, 836--846 (2014).
\bibitem{3} M.L. Bertotti, G. Modanese, 
Mathematical models describing the effects of different tax evasion behaviours,
submitted (2015).
\bibitem{4} M.L. Bertotti, G. Modanese,
From microscopic taxation and redistribution models to macroscopic income distributions,
{\it{Physica A}} \textbf{390}, 3782--3793 (2011).
\bibitem{5} M.L. Bertotti, G. Modanese,
Exploiting the flexibility of a family of models for taxation and redistribution,
{\it{Eur. Phys. J. B}} \textbf{85}, 261 (2012).
\bibitem{6} R. Levaggi, F. Menoncin, 
Tax audits, fines and optimal tax evasion in a dynamic context, 
{\it{Econ Lett}} 117, 318---321 (2012).
\bibitem{7} R. Levaggi, F. Menoncin, 
Dynamic tax evasion with audits based on conspicuous consumption, 
{\it{preprint}} (2015).
\bibitem{8} K.M. Bloomquist, M. Koehler, 
A large-scale agent-based model of taxpayer reporting compliance, 
{\it{J. Artif. Soc. Soc. Simul. }} 18, 20 (2015).
\bibitem{9} G. Zaklan, F. Westerhof, D. Stauffer,
Analysing tax evasion dynamics via the Ising Model,
{\it {J Econ Interact Coord}} {\textbf 4}, 1-14 (2009).
\bibitem{10} S. Hokamp, M. Pickhardt,
Income tax evasion in a society of heterogeneous agents: evidence from an agent-based model,
{\it{International Economic Journal}} \textbf{24}, 541--553 (2010).
\bibitem{11} S. Hokamp, G. Seibold, 
Tax compliance and public goods provision. An agent-based econophysics approach,
{\it{CEJEME}} {\bf 6}, 217-236 (2014).
\bibitem{12} N. Crokidakis,
A three-state kinetic agent-based model to analyse tax evasion dynamics,
{\it{Physica A}} \textbf{414}, 321-328 (2014).
\bibitem{13} F.W.S. Lima,
Tax evasion dynamics and non equilibrium Zaklan model with heterogeneous agents on square lattice,
{\it{Int. J. Mod. Phys. C}} \textbf{26}, 1550035 (8 pages) (2015).
\bibitem{14} K.M. Bloomquist,
Multi-agent based simulation of the deterrent effects of taxpayer audits, 
{\it {Proceedings. Annual Conference on Taxation and Minutes of the Annual Meeting of the National Tax Association}},
National Tax Association {\textbf 97}, 159-173 (2004). 
\bibitem{15} K.M. Bloomquist,
A Comparison of Agent-Based Models of Income Tax Evasion,
{\it{Social Science Computer Review}} {\textbf 24}, 411-425 (2006).
\bibitem{16} M.L. Bertotti, G. Modanese,
Microscopic models for welfare measures addressing a reduction of economic inequality,
{\it{Complexity}}, Article first published online, doi: 10.1002/cplx.21669 (2015).
\end{thebibliography}
\end{document}